\documentclass[conference]{IEEEtran}
\usepackage{amsmath,amssymb,bm}
\usepackage{graphicx}
\usepackage{epstopdf}
\usepackage{booktabs}
\usepackage{cite}
\usepackage{balance}
\usepackage{url}
\newcommand{\arxivurl}[1]{\url{#1}}

\makeatletter
\renewcommand{\subsection}{\@startsection{subsection}{2}{\z@}%
  {0.75ex plus 0.10ex minus 0.10ex}%
  {0.45ex plus 0.10ex}%
  {\normalfont\normalsize\itshape}}
\makeatother
\newcommand{\mS}{\mathcal{S}}
\newcommand{\mB}{\mathcal{B}}
\newcommand{\IA}{\textsc{IA-ZPA}}
\newcommand{\sg}{\operatorname{sg}}

\usepackage[font=small,skip=2pt]{caption}

\usepackage{enumitem}
\setlist{nosep}

\begin{document}
\title{Impedance-Aware Zonal Port Activation for Fluid Antenna Arrays}
\author{
	\IEEEauthorblockN{
		Yuanhui Wu\textsuperscript{1},
		Hao Jiang\textsuperscript{2},
		Zaichen Zhang\textsuperscript{3,4}
	}
	\IEEEauthorblockA{
		\textsuperscript{1}College of Artificial Intelligence, Nanjing University of Information Science and Technology, China\\
		\textsuperscript{2}School of Cyber Science and Engineering, Southeast University, China\\
		\textsuperscript{3}National Mobile Communications Research Laboratory, Southeast University, China\\
		\textsuperscript{4}Purple Mountain Laboratories, Nanjing, China\\[-1mm]
		Emails: 202412621447@nuist.edu.cn, jiang.hao@seu.edu.cn, zczhang@seu.edu.cn
	}
}
\maketitle

\begin{abstract}
Fluid antenna array (FAA) activation jointly determines the effective multi-user channel for precoding and the sparse physical aperture. Channel-oriented selection can concentrate high-gain ports and erode aperture quality, whereas geometry-oriented selection does not adapt to instantaneous channel state information (CSI). This paper formulates finite-port FAA activation as a rate--aperture--feasibility problem under an exact RF-chain budget. We propose impedance-aware zonal port activation (IA-ZPA), which couples compact CSI-conditioned port scoring with a checkerboard feasibility projection and inference-time mutual-impedance-aware selection. The learned scorer ranks ports, while the deterministic rule fixes the active aperture; a separate current-domain RZF backend then evaluates source-drive feasibility. Under a common induced-EMF protocol, IA-ZPA attains the largest constrained rate among the methods satisfying the prescribed mean-PSLL target with a substantially lower decision time than greedy selection.
\end{abstract}

\section{Introduction}
Beamforming exploits spatial degrees of freedom for array gain and multiuser interference control. In conventional arrays, the element coordinates, and thus the aperture sampling geometry, are fixed after deployment. The aperture cannot then respond to changes in the channel, user geometry, or radiation requirement.

Fluid antenna systems (FASs)~\cite{wong2020_fas_limits,Wong2021FAS,Zhang2026FBLTWC} allow a limited number of RF chains to select from prescribed radiating ports. Candidate ports experience spatially correlated but nonidentical fading. Selecting a port therefore selects a favorable point on the channel-gain envelope without assigning an RF chain to every candidate location. This observation has led to work on port-domain statistics and CSI reconstruction~\cite{Zhang2026FBL,Zhang2026SpatialCorr,New2025ChannelRecon,Zhang2026GeometryRecon}, as well as port selection and CSI acquisition with limited observations or processing~\cite{Zhang2025CSI,Chai2022PortSelection,Zou2024Online,Efrem2024FluidMIMO}. Learning-based reconstruction and selection, together with hardware-oriented beamforming, have also been studied for practical activation~\cite{Liang2026ChannelRecon,Eskandari2025CGAN,JSAC,Wu2026LAMP,Xu2026Practical}.

Fluid antenna arrays (FAA) \cite{Zhang2026FiniteAperture} take this idea beyond the selection of one favorable fading realization: its port mask configures a reconfigurable multiport aperture and creates \emph{geometry diversity}. The mask changes the effective multiuser channel, the sampled aperture, and the array manifold simultaneously. Finite-aperture FAA analysis, planar FAA modeling, and electromagnetic-aware FAA design establish the basis for treating these effects jointly~\cite{Zhang2026FiniteAperture,Chen2026Nearfield,Zhang2024JointPortBF,Zhang2026PlanarFAA,Craeye2011,Zhang2026EMAware}. A channel-only rule may cluster active ports, while a fixed geometry-only mask cannot follow instantaneous CSI. Recent work has examined array-level characterization, flexible beam control, sidelobe-oriented configuration, real-time activation, and constraint-aware design~\cite{Xu2026Flexible,Buck2023,Liang2026PSLL,Ye2024EMExposure,Wu2026BLPA}.

These coupled effects lead to an online finite-port selection problem under channel, aperture, RF-chain, and source-drive constraints. The mask must support multiuser precoding and a distributed sparse aperture within the port budget, while avoiding the added drive burden caused by mutual coupling between nearby ports. Since a full electromagnetic or circuit-constrained optimization is impractical for each channel realization, IA-ZPA combines CSI-conditioned scoring with checkerboard projection to enforce exact zone-wise quotas. An induced-EMF kernel penalizes strongly coupled choices during inference, and a common current-domain RZF backend evaluates the resulting masks under accepted-power, current, and source-voltage limits.

The contributions are as follows.
\begin{itemize}
\item A channel--aperture--feasibility formulation in which one mask determines the selected channel used by RZF, the equal-amplitude aperture diagnostic, and the current-domain check.
\item \IA{}, which combines compact CSI-conditioned scoring, checkerboard projection, and inference-time mutual-impedance-aware selection. The projection satisfies every per-zone quota exactly, so the RF-chain budget and aperture coverage do not depend on the learned scores alone.
\item A common induced-EMF evaluation protocol that places all selectors under the same current-domain source-drive constraints and reports equal-amplitude PSLL together with CSI-conditioned online mask-decision time.
\end{itemize}

The remainder of this paper is organized as follows. Section~\ref{sec:system} presents the system model and evaluation procedure. Section~\ref{sec:method} describes IA-ZPA, Section~\ref{sec:results} reports the results, and Section~\ref{sec:conclusion} concludes the paper.

\noindent\textit{Notation:} Scalars, vectors, and matrices are denoted by italic letters, bold lowercase letters, and bold uppercase letters, respectively. $\mathbb C^{m\times n}$ denotes the set of $m\times n$ complex matrices, and $\mathbf I$ denotes an identity matrix of conformable dimension. The operators $(\cdot)^T$, $(\cdot)^H$, $\operatorname{tr}(\cdot)$, $\Re\{\cdot\}$, $|\cdot|$, $\|\cdot\|_0$, and $\|\cdot\|_F$ denote transpose, Hermitian transpose, trace, real part, absolute value, number of nonzero entries, and Frobenius norm, respectively.

\section{System Model and Problem Formulation}\label{sec:system}
\subsection{Finite-Port Downlink and Two-Stage Evaluation}
Consider $N$ candidate ports at wavelength-normalized coordinates $\{\mathbf p_n\}_{n=1}^{N}$ and $K$ single-antenna users. The binary mask and active set obey
\begin{equation}
 \mathbf a\in\{0,1\}^{N},~\mS=\{n:a_n=1\},~\|\mathbf a\|_0=M.
 \label{eq:mask}
\end{equation}
For user $k$, a UMi-inspired model draws the LoS state and generates five clusters with 12 rays each. The path gains $\alpha_{k,\ell}$ include distance-dependent path loss, shadowing, blockage, and a Rician LoS component when present. These paths produce the narrowband finite-port response $\bar h_{k,n}=\sum_{\ell=1}^{L_k}\alpha_{k,\ell}\exp(j2\pi\mathbf p_n^T\mathbf u_{k,\ell}^{\rm q})$, where $\mathbf u_{k,\ell}^{\rm q}$ lies on the visible subset of a direction-cosine grid $\mathcal U$. We normalize this response over the full candidate aperture, $h_{k,n}=\bar h_{k,n}/\sqrt{N^{-1}\sum_m|\bar h_{k,m}|^2}$, and collect the entries selected by \eqref{eq:mask} in $\mathbf H_{\mS}\in\mathbb C^{K\times M}$. Hence, the reported normalized-SNR experiments compare spatial channel structure; distance controls the pre-normalization path statistics rather than the final user-to-user SNR spread.

One activation changes both the reduced multiuser channel and the sampled aperture. We therefore use two linked evaluations. During training, the unit-Frobenius-norm RZF map $\widetilde{\mathbf P}_{\mS}\propto\mathbf H_{\mS}^{H}(\mathbf H_{\mS}\mathbf H_{\mS}^{H}+\lambda_{\rm RZF}\mathbf I)^{-1}$, with $\|\widetilde{\mathbf P}_{\mS}\|_F^2=1$~\cite{Wagner2012RZF}, measures the CSI value of a candidate mask. With $\mathbf G_{\mS}=\mathbf H_{\mS}\widetilde{\mathbf P}_{\mS}$ and reference SNR $\rho$, the resulting proxy is
\begin{equation}
 R_{\rm sel}(\mathbf a)=\sum_{k=1}^{K}\log_2\!\left(1+\frac{\rho|[\mathbf G_{\mS}]_{k,k}|^2}{1+\rho\sum_{j\ne k}|[\mathbf G_{\mS}]_{k,j}|^2}\right).
 \label{eq:selection-score}
\end{equation}
During training, $\rho_{\rm dB}=10+\Delta$ with $\Delta\sim\mathcal U[-5,5]$ dB; validation and the zero-shot diagnostics use $\rho=10$ dB. This proxy supplies the channel-dependent learning signal. After activation, Section~\ref{sec:method} evaluates every selected subarray using a common current-domain RZF backend under source-drive constraints.

The equal-amplitude broadside array factor is
\begin{equation}
 A_{\mS}(u,v)=\frac{1}{M}\sum_{n\in\mS}
 \exp\!\left(j2\pi\mathbf p_n^T[u,v]^T\right).
 \label{eq:af}
\end{equation}
The PSLL is the maximum of $20\log_{10}|A_{\mS}(u,v)|$ over the visible grid outside the nominal first-null broadside guard rectangle determined by the full aperture. It complements $R_{\rm sel}$ by evaluating the geometry of the sparse aperture under a common equal-amplitude excitation.

\subsection{Activation Problem}
For every channel realization, the selector maps full-aperture CSI to an $M$-port mask. The desired mask preserves multiuser separability through $R_{\rm sel}(\mathbf a)$, maintains a distributed aperture through the PSLL in \eqref{eq:af}, and admits an electrically efficient current-domain realization. IA-ZPA realizes this sequence by learning CSI scores, projecting them onto a fixed aperture layout, and applying coupling-aware tie breaking during deployment.

\section{Proposed IA-ZPA Method}\label{sec:method}
\subsection{CSI Scoring and Exact Zone-wise Activation}
IA-ZPA separates CSI preference from aperture feasibility. Each candidate port is represented by its coordinates, the log powers of the across-user complex-field mean $\mu_n$ and standard deviation $\sigma_n$, the log coherence proxy $q_n=|\mu_n|^2/(1+|\sigma_n|^2)$, their phase difference, and the log mean and standard deviation of the per-user powers. A compact CNN maps the standardized feature tensor to a CSI-conditioned score $z_n$.

IA-ZPA first enforces aperture coverage. Let $a_n\in\{0,1\}$ select port $n$, and partition the aperture into zones $\mB_b$ with exactly $m_b$ active ports per zone:
\begin{equation}
 \sum_{n\in\mB_b}a_n=m_b,~\sum_bm_b=M.
 \label{eq:zone}
\end{equation}
The zone quota fixes the RF-chain budget and distributes the selected ports across the aperture. It turns the CNN score field into a mask with a prescribed spatial footprint.

\subsection{Learning the CSI Score Field}
The score field is learned with a differentiable, zone-wise relaxation. For zone $b$, Gumbel perturbations $g_n$ define the soft top-$m_b$ relaxation in \eqref{eq:softmask}~\cite{Jang2017GumbelSoftmax}:
\begin{equation}
 a_{b,n}^{\rm soft}=\left[m_b\,
 \mathrm{softmax}\!\left((z_n+g_n)/\tau\right)\right]_{[0,1]},
 \label{eq:softmask}
\end{equation}
where $[\cdot]_{[0,1]}$ clips the value to $[0,1]$ and $\tau$ is the relaxation temperature. The forward pass uses the zone-wise hard Top-$m_b$ mask $\mathbf a^{\rm hard}$, while the straight-through form~\cite{Bengio2013STE}
\begin{equation}
 \mathbf a^{\rm tr}=\mathbf a^{\rm hard}+\mathbf a^{\rm soft}-\sg(\mathbf a^{\rm soft})
 \label{eq:st}
\end{equation}
passes gradients through the soft mask. The training objective is
\begin{equation}
 \mathcal L=-\lambda_R R_{\rm sel}+\lambda_{\rm sl}\mathcal L_{\rm sl}
 +\lambda_{\rm card}\mathcal L_{\rm card}+\lambda_{\rm bin}\mathcal L_{\rm bin}.
 \label{eq:loss}
\end{equation}
Let $\mathcal G_{\rm sl}$ contain the training-grid points outside the same first-null guard used for PSLL. With the soft-mask array factor $A_{\mathbf a^{\rm tr}}$, the smooth sidelobe penalty is
\begin{equation}
\begin{aligned}
 \mathcal L_{\rm sl}&=\kappa^{-1}\log\!\left[\frac{1}{|\mathcal G_{\rm sl}|}
 \sum_{(u,v)\in\mathcal G_{\rm sl}}\right.\\[-0.2ex]
 &\left.~\times\exp\!\left\{\kappa\left(20\log_{10}|A_{\mathbf a^{\rm tr}}(u,v)|-\beta_{\rm sl}\right)\right\}\right].
\end{aligned}
 \label{eq:sidelobe-loss}
\end{equation}
The terms $\mathcal L_{\rm card}$ and $\mathcal L_{\rm bin}$ stabilize the soft relaxation. Training therefore shapes a CSI score field that balances the effective channel and sparse-aperture response while preserving the zone quota in the forward pass.

\subsection{Inference-Time Coupling-Aware Projection}
For parallel half-wave dipoles, the induced-EMF mutual impedance at separation $d$ is $Z^{\rm ind}(d)=\eta_0[2F(kd)-F(k(r+\ell))-F(kd^2/(r+\ell))]/(4\pi)$, where $F(x)=\operatorname{Ci}(x)-j\operatorname{Si}(x)$, $k=2\pi/\lambda$, $\ell=\lambda/2$, $r=\sqrt{d^2+\ell^2}$, and $\eta_0=376.73~\Omega$. We set $Z_{\rm self}=73.1+j42.5~\Omega$ and use the dimensionless pair kernel
\begin{equation}
 C_{mn}=\left|Z_{mn}^{\rm ind}/Z_{\rm self}\right|^2,~C_{nn}=0,
 \label{eq:kernel}
\end{equation}
to summarize pairwise electrical interaction during discrete selection. Index port $n$ by its grid coordinates $(i_n,j_n)$ and define the parity-$c$ candidate set in zone $b$ as $\mB_b^{(c)}=\{n\in\mB_b:(i_n+j_n)\bmod2=c\}$, $c\in\{0,1\}$. Zones are visited in raster order. For each parity, the $t$th choice in zone $b$ is
\begin{equation}
\begin{aligned}
 n_{b,t}^{(c)}&=\underset{n\in\mB_b^{(c)}\setminus\mS_{<b,t}}{\arg\max}
 \left\{\frac{z_n-\mu_b}{\sigma_b+10^{-9}}-\omega\sum_{m\in\mS_{<b,t}}C_{nm}\right\},\\[-0.2ex]
 &\hspace{7.5em}\omega=1,
\end{aligned}
 \label{eq:projection}
\end{equation}
where $\mu_b$ and $\sigma_b$ are the mean and standard deviation of the raw CNN scores in zone $b$, and $\mS_{<b,t}$ contains all earlier choices. After satisfying the $m_b$ quota in every zone, the two masks are compared by $U_c=\sum_{b,t}s_{n_{b,t}^{(c)}}-\omega M\bar C(\mS^{(c)})$, where $s_n$ is the bracketed score in \eqref{eq:projection} and $\bar C$ is the average $C_{mn}$ over distinct ordered selected-port pairs. The mask with larger $U_c$ is deployed. This projection enforces \eqref{eq:zone}, preserves the checkerboard layout, and uses mutual-impedance information at the discrete-choice stage.

\subsection{Current-Domain Precoding and Online Complexity}
After the mask is fixed, every selector is evaluated by the same current-domain RZF calculation. The selected-port impedance matrix has off-diagonal entries $Z_{mn}^{\rm ind}$ and diagonal entries $Z_{\rm self}+R_\ell$, where $R_\ell=1~\Omega$. Let $\mathbf R_{\rm acc}=(\mathbf Z_{\mS}+\mathbf Z_{\mS}^{H})/2$, and let $R_s$ be the source impedance. With RMS phasors, $\operatorname{tr}(\mathbf W^H\mathbf R_{\rm acc}\mathbf W)$ is the accepted power, and the source-voltage metric is
\begin{equation}
 \mathbf Q_v=(\mathbf Z_{\mS}+R_s\mathbf I)^H(\mathbf Z_{\mS}+R_s\mathbf I).
 \label{eq:qv}
\end{equation}
The calculation uses the same unscaled selected channel $\mathbf H_{\mS}$ as the training proxy. With $\bar r=\operatorname{tr}(\mathbf R_{\rm acc})/M$ and $\bar q=\operatorname{tr}(\mathbf Q_v)/M$, the metric and RZF solution are
\begin{align}
 \boldsymbol\Xi&=\frac{\mathbf R_{\rm acc}}{\bar r}+\eta_I\mathbf I+\eta_V\frac{\mathbf Q_v}{\bar q}, \label{eq:metric}\\
 \mathbf B&=\boldsymbol\Xi^{-1}\mathbf H_{\mS}^{H},~
 \mathbf W_0=\mathbf B(\mathbf H_{\mS}\mathbf B+\lambda_{\rm RZF}\mathbf I)^{-1}.
 \label{eq:current-rzf}
\end{align}
The metric in \eqref{eq:metric} weights accepted-power and source-voltage usage before a common scaling. We scale $\mathbf W_0$ by the largest $\gamma\leq1$ satisfying
\begin{equation}
\begin{aligned}
 \mathrm{tr}(\mathbf W^H\mathbf R_{\rm acc}\mathbf W)&\leq P_{\rm acc},\\
 \|\mathbf W\|_F^2&\leq I_{\max}^2,\\
 \mathrm{tr}(\mathbf W^H\mathbf Q_v\mathbf W)&\leq V_{\max}^2.
\end{aligned}
\label{eq:limits}
\end{equation}
For the isolated-port 10-dB reference, we set
\begin{equation}
 N_0=\frac{\operatorname{median}_{k,n}|h_{k,n}|^2P_{\rm acc}}
 {10\bigl(\Re\{Z_{\rm self}\}+R_\ell\bigr)}.
 \label{eq:noise}
\end{equation}
With $\mathbf G_{\rm EM}=\mathbf H_{\mS}\mathbf W$, the constrained sum rate is $R_{\rm EM}=\sum_k\log_2\!\left(1+|[\mathbf G_{\rm EM}]_{k,k}|^2/(N_0+\sum_{j\ne k}|[\mathbf G_{\rm EM}]_{k,j}|^2)\right)$. The scaling factor $\gamma\in(0,1]$ summarizes the drive margin: a larger value corresponds to less common backoff. Together, $R_{\rm EM}$ and $\gamma$ quantify the communication and electrical consequences of the selected mask under one backend.

Online IA-ZPA comprises one scorer pass and zone-wise coupling-aware selection; coupling updates require at most $\mathcal O(NM)$ kernel accesses. The shared current-domain backend costs $\mathcal O(M^3+M^2K+K^3)$. Table~\ref{tab:comparison} reports the channel-dependent mask-decision time, while the common backend is kept separate from all selector timings.

\begin{figure*}[!t]
\centering
\includegraphics[width=0.94\textwidth]{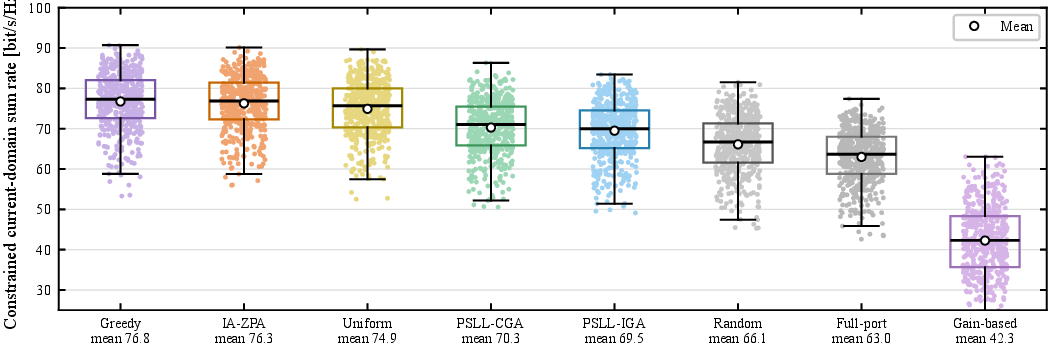}
\caption{Constrained current-domain sum-rate distributions over 500 common 256--64 test channels under the same induced-EMF model and source-drive limits. White circles denote means; IA-ZPA averages three clean checkpoints.}
\label{fig:em-ratebox}
\end{figure*}

\section{Simulation Setup and Results}\label{sec:results}
\subsection{Setup and Baselines}
We use a $16\times16$ candidate grid over a $5\lambda\times2.5\lambda$ aperture ($N=256$), activate $M=64$ ports for $K=16$ users, and impose one active port in each of $8\times8$ zones. The finite-port channels use a $50\times25$ visible direction-cosine codebook and user distances in $[50,150]$ m. The split contains 1200/200/500 training/validation/test channels; every baseline uses the same 500 unscaled test channels, and IA-ZPA averages three independently trained clean checkpoints. PSLL is recomputed on a $301\times301$ direction-cosine grid. All online selectors receive precomputed full candidate-port CSI; timing begins after this CSI-derived input is available and excludes channel acquisition.

The scorer has eight input channels, two $3\times3$ Conv--GN--ReLU stages of width 32, and a $1\times1$ head. We use $\tau=0.6$, $\lambda_{\rm RZF}=0.1$, $(\lambda_R,\lambda_{\rm sl},\lambda_{\rm card},\lambda_{\rm bin})=(0.14,1.2,10^{-2},10^{-3})$, a $51\times51$ sidelobe-loss grid with the preselected $\beta_{\rm sl}=-13.5$ dB and $\kappa=80$, and $(\eta_I,\eta_V)=(0.05,0.02)$. The same preselected $-13.5$-dB threshold is used only to label the mean-PSLL target in Table~\ref{tab:comparison}; it is not chosen from the reported test results. The proxy uses a 10-dB reference SNR with a uniform $\pm5$-dB training jitter. The induced-EMF check sets $P_{\rm acc}=1$ W, $I_{\max}^2=0.03621$, $V_{\max}^2=443.4$, $R_s=50~\Omega$, and 1-$\Omega$ loss resistance; noise is calibrated to a 10-dB isolated-port reference. The comparison methods are:
\begin{itemize}
\setlength\itemsep{0pt}
\setlength\parskip{0pt}
\item \textbf{IA-ZPA (proposed)}: applies CNN scoring, checkerboard zone projection, and inference-time mutual-impedance-aware selection.
\item \textbf{Full-port RZF}: activates all 256 ports under the same aggregate drive budgets; it is a dense baseline, not a rate upper bound.
\item \textbf{Uniform}: uses a fixed uniformly distributed mask.
\item \textbf{PSLL-CGA}: uses an offline geometry-only PSLL mask from the compact genetic algorithm.
\item \textbf{PSLL-IGA}: uses an offline geometry-only PSLL mask from the iterative genetic algorithm.
\item \textbf{Random}: resamples active ports without CSI.
\item \textbf{Gain-based}: ranks aggregate channel power.
\item \textbf{Greedy}: sequentially maximizes the normalized RZF-rate proxy over the candidate set.
\end{itemize}

\begin{table}[t]
\centering
\caption{Common 500-channel current-domain comparison under identical channel, induced-EMF, noise, and aggregate source-drive budgets.}
\label{tab:comparison}
\small
\renewcommand{\arraystretch}{1.03}
\setlength{\tabcolsep}{0pt}
\begin{tabular*}{\columnwidth}{@{\extracolsep{\fill}}lrrrrr@{}}
\toprule
Method & $R_{\rm EM}$ & $\gamma$ & PSLL & Mean target & Decision\\
\midrule
\multicolumn{6}{@{}l}{\textit{Online adaptive selectors}}\\[-0.2ex]
Greedy & \textbf{76.75} & \textbf{0.1742} & -10.24 & No & 199.74\\
IA-ZPA & \textbf{76.28}$^{\dagger}$ & 0.1723 & -13.66 & Yes & 2.73\\
Random & 66.12 & 0.1372 & -10.47 & No & 0.023\\
Gain-based & 42.26 & 0.0782 & -5.64 & No & \textbf{0.020}\\
\multicolumn{6}{@{}l}{\textit{Fixed or offline masks}}\\[-0.2ex]
Uniform & 74.92 & 0.1673 & -10.26 & No & --\\
PSLL-CGA & 70.29 & 0.1507 & \textbf{-17.89} & Yes & --\\
PSLL-IGA & 69.50 & 0.1481 & -16.45 & Yes & --\\
Full-port RZF & 62.99 & 0.1271 & -13.15 & No & --\\
\bottomrule
\end{tabular*}
\vspace{2.5pt}
\raggedright\footnotesize\textit{Note:} $R_{\rm EM}$ is in bit/s/Hz, PSLL in dB, and Decision in ms. ``Mean target'' denotes the preselected $\overline{\mathrm{PSLL}}\leq-13.5$ dB criterion; $^{\dagger}$ marks its highest $R_{\rm EM}$. Larger $\gamma$ means less backoff. Decision is the three-repeat median over 500 channels; one checkpoint is used per IA-ZPA call and results are pooled. It includes one scorer pass and mask construction, but excludes CSI acquisition and the common RZF backend.
\end{table}

\subsection{Rate--PSLL--Time Tradeoff}
Table~\ref{tab:comparison} and Fig.~\ref{fig:em-ratebox} use one current-domain backend, so $R_{\rm EM}$ is directly comparable. Among the methods meeting the mean PSLL target, IA-ZPA attains the largest constrained rate. Greedy reaches 76.75 bit/s/Hz, 0.47 bit/s/Hz above IA-ZPA, but its mean PSLL is $-10.24$ dB and its decision time is 199.74 ms. IA-ZPA raises $R_{\rm EM}$ by 1.36 bit/s/Hz over Uniform while reducing PSLL by 3.40 dB. The two static PSLL masks achieve the lowest PSLL values, whereas Random and Gain-based provide low-cost decisions with weaker rate--PSLL tradeoffs. Full-port RZF reaches the voltage limit before its accepted-power and current budgets ($0.234<1$ W and $0.00831<0.03621$), which explains its lower constrained rate under the common drive budget.

\subsection{Module Attribution}
Table~\ref{tab:ablation} holds the training objective and 500 channels fixed. ``CNN score only'' retains the zone quota but removes both the checkerboard restriction and the kernel penalty; ``CNN + checkerboard'' adds the two-lattice feasibility projection; and Full IA-ZPA additionally applies the inference-time mutual-impedance kernel.

\begin{table}[t]
\centering
\caption{Inference-module attribution under the common 500-channel induced-EMF current-domain protocol. All learned variants share the same clean checkpoints and training objective; smaller $\bar C$ indicates weaker average pairwise coupling.}
\label{tab:ablation}
\small
\renewcommand{\arraystretch}{1.10}
\setlength{\tabcolsep}{0pt}
\begin{tabular*}{\columnwidth}{@{\extracolsep{\fill}}lrrrr@{}}
\toprule
Variant & $R_{\rm EM}$ & PSLL & $\gamma$ & $\bar C$\\
\midrule
Uniform-zone & 74.92 & -10.26 & 0.1673 & \textbf{0.02543}\\
CNN score only & 75.34 & \textbf{-13.67} & 0.1688 & 0.02786\\
CNN + checkerboard & 76.17 & -13.61 & 0.1719 & 0.02731\\
Full IA-ZPA & \textbf{76.28} & -13.66 & \textbf{0.1723} & 0.02718\\
\bottomrule
\end{tabular*}
\end{table}

Relative to Uniform-zone, CNN scoring lowers PSLL by 3.41 dB while increasing $R_{\rm EM}$ by 0.42 bit/s/Hz. Adding the checkerboard projection raises $R_{\rm EM}$ by a further 0.82 bit/s/Hz, increases $\gamma$ from 0.1688 to 0.1719, and imposes a $\lambda/3$ minimum selected-port spacing rather than the $\lambda/6$ spacing observed for score-only selection. The kernel lowers the mean pair measure from 0.02731 to 0.02718 (0.46\%) and then changes $R_{\rm EM}$ by 0.11 bit/s/Hz and $\gamma$ by 0.0004. Thus, CNN scoring and projection account for most of the observed rate--aperture--feasibility tradeoff, whereas the inference-time kernel is a small electrical refinement rather than the dominant source of gain.

\subsection{Aperture-Pattern Diagnostic}
\begin{figure}[!t]
\centering
\includegraphics[width=\columnwidth]{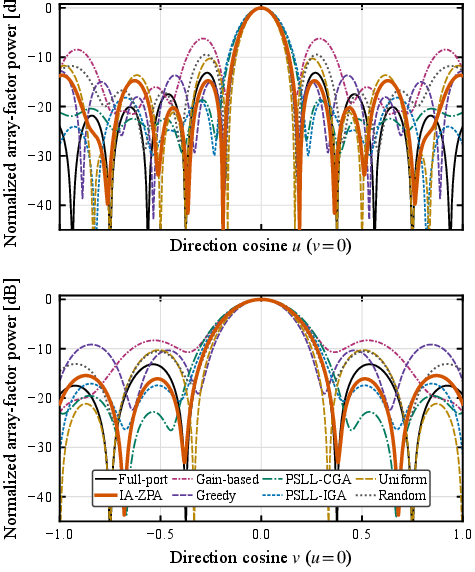}
\caption{Equal-amplitude normalized array-factor cuts for a 256--64 test realization selected by a fixed above-median-rate and mean-PSLL-proximity rule; every method uses this same channel. The legend is inside the lower panel.}
\label{fig:cuts}
\end{figure}

Fig.~\ref{fig:cuts} shows the $u$- and $v$-axis cuts of the equal-amplitude array factor for one held-out realization. All curves are normalized to the same broadside peak; the off-axis structure therefore reveals the aperture effect of the selected masks.

The selected masks retain the broadside main lobe, but their off-axis behavior differs markedly. In the $u$ cut, Gain-based produces the largest visible sidelobes, while the IA-ZPA curve suppresses these lobes and retains deep neighboring nulls. In the $v$ cut, IA-ZPA also remains below Gain-based and Greedy around the first sidelobe region. The PSLL-CGA and PSLL-IGA masks attenuate several axial lobes further, consistent with their geometry-first objective. Since Table~\ref{tab:comparison} reports PSLL on the full two-dimensional grid, the cuts provide an interpretable directional view of, rather than a replacement for, that aggregate metric.

\subsection{Zero-Shot Port and User Sweeps}
\begin{figure}[!t]
\centering
\includegraphics[width=\columnwidth]{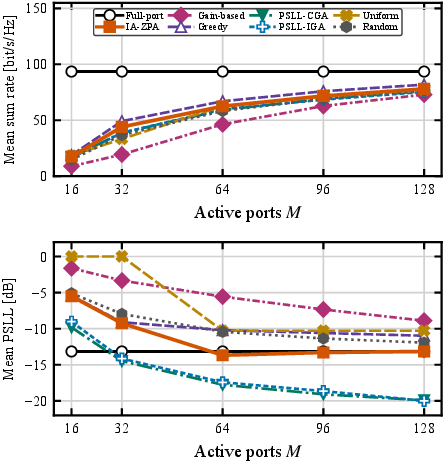}
\caption{Active-port sweep of all eight selectors at $K=16$ under the normalized RZF-rate proxy: mean sum rate (top) and equal-amplitude PSLL (bottom). Full-port is fixed at $N=256$.}
\label{fig:active-sweep}
\end{figure}

\begin{figure}[!t]
\centering
\includegraphics[width=\columnwidth]{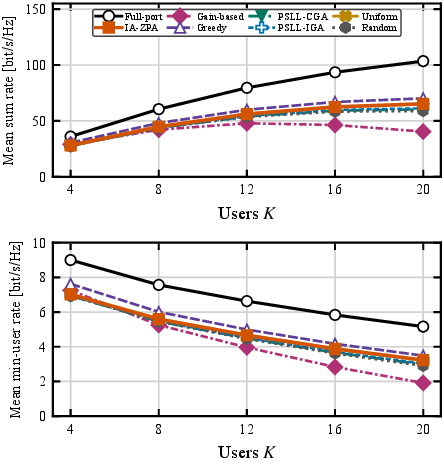}
\caption{User-count sweep of all eight selectors at $M=64$ under the normalized RZF-rate proxy: mean sum rate (top) and minimum-user rate (bottom). Full-port is fixed at $N=256$. This is separate from current-domain evaluation.}
\label{fig:user-sweep}
\end{figure}

Figs.~\ref{fig:active-sweep} and~\ref{fig:user-sweep} examine zero-shot load sensitivity of the selector trained at $(M,K)=(64,16)$; each point averages 300 channels. The active-port sweep varies $M=16$--128 at $K=16$, and the user sweep varies $K=4$--20 at $M=64$. We use the normalized RZF proxy here to isolate the effect of changing the activation or multiplexing load. Sum rate increases with more active sparse ports or served streams, whereas the minimum-user rate falls as a fixed aperture and drive budget are shared among more users.

\subsection{Practical Interpretation}
The results separate three roles: Table~\ref{tab:comparison} and Fig.~\ref{fig:em-ratebox} quantify constrained current-domain rate, PSLL, and decision time; the sweeps isolate selector load sensitivity; and Fig.~\ref{fig:cuts} exposes the aperture patterns behind the PSLL comparison. The coupling kernel guides the discrete choice, while $\mathbf R_{\rm acc}$ and $\mathbf Q_v$ determine the subsequent electrical backoff.

Under the fixed aggregate drive budget, dense RZF reaches the array-voltage limit before exhausting accepted power or total current. A distributed 64-port mask can therefore achieve a larger $R_{\rm EM}$ in this induced-EMF model. The calculation uses the impedance submatrix of the selected ports; incorporating terminated inactive ports and EEP-derived fields would extend the same framework from circuit-constrained rate evaluation to realized-radiation assessment.

\section{Conclusion}\label{sec:conclusion}
We have presented IA-ZPA for sparse FAA activation with CSI-conditioned scoring, checkerboard projection, inference-time impedance-aware selection, and constrained current-domain RZF. In the common induced-EMF evaluation, IA-ZPA attains the largest constrained rate among the methods meeting the mean-PSLL target, while reducing the online decision time from 199.74 ms for Greedy to 2.73 ms. The aperture cuts further show how the learned CSI preference is converted into a distributed mask with controlled off-axis structure. IA-ZPA achieves an effective tradeoff among constrained sum rate, sidelobe control, and online decision latency. Extending the evaluation to terminated inactive ports and EEP-based realized radiation is a relevant next step.

\balance


\begin{thebibliography}{99}
\bibitem{wong2020_fas_limits}K.-K. Wong, \emph{et al.}, ``Performance limits of fluid antenna systems,'' \emph{IEEE Commun. Lett.}, vol.~24, no.~11, pp.~2469--2472, Nov.~2020.
\bibitem{Wong2021FAS}K.-K. Wong, \emph{et al.}, ``Fluid antenna systems,'' \emph{IEEE Trans. Wireless Commun.}, vol.~20, no.~3, pp.~1950--1962, Mar.~2021.
\bibitem{Zhang2026FBLTWC}Z. Zhang, \emph{et al.}, `Finite-blocklength fluid antenna systems,'' \emph{IEEE Trans. Wireless Commun.}, early access, 2026, doi: 10.1109/TWC.2026.3723456.
\bibitem{Zhang2026FBL}Z. Zhang, \emph{et al.}, ``Finite-blocklength fluid antenna systems with spatial block-correlation channel model,'' \emph{IEEE Wireless Commun. Lett.}, vol.~15, pp.~1911--1915, 2026.
\bibitem{Zhang2026SpatialCorr}Z. Zhang, \emph{et al.}, ``Beyond covariance: Generative spatial correlation modeling and channel interpolation for fluid antenna systems,'' \emph{arXiv preprint}, 2026. [Online]. Available: \arxivurl{https://arxiv.org/abs/2604.16639}
\bibitem{New2025ChannelRecon}W. K. New, \emph{et al.}, ``Channel estimation and reconstruction in fluid antenna system: Oversampling is essential,'' \emph{IEEE Trans. Wireless Commun.}, vol.~24, no.~1, pp.~309--322, Jan.~2025, doi: 10.1109/TWC.2024.3491507.
\bibitem{Zhang2026GeometryRecon}Z. Zhang, \emph{et al.},  ``Geometry-structured channel reconstruction for conventional and fluid antenna systems: Bayesian inference and fundamental limits,'' \emph{arXiv preprint}, 2026. [Online]. Available: \arxivurl{https://arxiv.org/abs/2606.04001}
\bibitem{Zhang2025CSI}Z. Zhang, \emph{et al.}, ``Low-complexity CSI acquisition exploiting geographical diversity in fluid antenna system,'' in \emph{Proc. IEEE Global Commun. Conf. Workshops (GC Wkshps)}, Dec.~2025, pp.~308--313.
\bibitem{Chai2022PortSelection}Z. Chai, \emph{et al.}, ``Port selection for fluid antenna systems,'' \emph{IEEE Commun. Lett.}, vol.~26, no.~5, pp.~1180--1184, May~2022.
\bibitem{Zou2024Online}J. Zou, S. Sun, and C. Wang, ``Online learning-induced port selection for fluid antenna in dynamic channel environment,'' \emph{IEEE Wireless Commun. Lett.}, vol.~13, no.~2, pp.~313--317, Feb.~2024.
\bibitem{Efrem2024FluidMIMO}Z. Zhang, \emph{et al.}, ``Cram\'er--Rao bounds for activity detection in conventional and fluid antenna systems,'' \emph{IEEE Wireless Commun. Lett.}, vol. 15, pp. 3059--3063, 2026.
\bibitem{Liang2026ChannelRecon}H. Liang, \emph{et al.}, ``Neural networks-enabled channel reconstruction for fluid antenna systems: A data-driven approach,'' in \emph{Proc. IEEE Wireless Commun. Netw. Conf. Workshops (WCNCW)}, 2026, pp.~1--6.
\bibitem{Eskandari2025CGAN}Z. Zhang, \emph{et al.}, ``On fundamental limits of slow-fluid antenna multiple access for unsourced random access,'' \emph{IEEE Wireless Commun. Lett.}, vol. 14, no. 11, pp. 3455--3459, 2025.
\bibitem{JSAC}
Z. Zhang, K.-K. Wong, J. Dang, Z. Zhang, and C.-B. Chae, ``On fundamental limits for fluid antenna-assisted integrated sensing and communications for unsourced random access,'' \emph{IEEE J. Sel. Areas Commun.}, vol. 44, pp. 136--149, 2026.
\bibitem{Wu2026LAMP}Y. Wu, \emph{et al.},  ``Learned-approximate message passing under Karhunen--Lo\`eve modeling for fluid antenna systems,'' \emph{IEEE Wireless Commun. Lett.}, vol.~15, pp.~2719--2723, 2026.
\bibitem{Xu2026Practical}S. Xu, \emph{et al.}, ``Toward practical fluid antenna systems: Co-optimizing hardware and software for port selection and beamforming,'' \emph{IEEE Trans. Wireless Commun.}, vol.~25, pp.~8341--8353, 2026.
\bibitem{Zhang2026FiniteAperture}Z. Zhang, \emph{et al.}, ``Finite-aperture fluid antenna array design: Analysis and algorithm,'' \emph{IEEE Wireless Commun. Lett.}, vol.~15, pp.~3199--3203, 2026.
\bibitem{Chen2026Nearfield}S. Chen, \emph{et al.}, ``Near-field beamforming and port selection for fluid antenna systems,'' \emph{IEEE Commun. Lett.}, vol.~30, pp.~2248--2252, 2026.
\bibitem{Zhang2024JointPortBF}Z. Zhang, \emph{et al.}, ``Joint activity detection and channel estimation for fluid antenna system exploiting geographical and angular information,'' \emph{IEEE J. Sel. Topics Signal Process.}, vol. 20, no. 3, pp. 354--370, 2026.
\bibitem{Zhang2026PlanarFAA}Z. Zhang, \emph{et al.}, ``Finite-aperture planar fluid antenna array,'' \emph{arXiv preprint}, 2026. [Online]. Available: \arxivurl{https://arxiv.org/abs/2605.22040}
\bibitem{Craeye2011}C. Craeye and D. Gonz\'alez-Ovejero, ``A review on array mutual coupling analysis,'' \emph{Radio Sci.}, vol.~46, RS2012, 2011.
\bibitem{Zhang2026EMAware}Z. Zhang, \emph{et al.}, ``Electromagnetic-aware fluid antenna array,'' \emph{arXiv preprint}, 2026. [Online]. Available: \arxivurl{https://arxiv.org/abs/2607.21375}
\bibitem{Xu2026Flexible}J. Xu, \emph{et al.}, ``Fluid antenna-enhanced flexible beamforming,'' in \emph{Proc. IEEE Wireless Commun. Netw. Conf. Workshops (WCNCW)}, 2026, pp.~1--6.
\bibitem{Buck2023}D. Buck, \emph{et al.}, ``Measuring array mutual impedances using embedded element patterns,'' \emph{IEEE Trans. Antennas Propag.}, vol.~71, no.~1, pp.~606--611, Jan.~2023.
\bibitem{Liang2026PSLL}H. Liang, \emph{et al.}, ``Peak sidelobe suppression in planar fluid antenna array,'' \emph{arXiv preprint}, 2026. [Online]. Available: \arxivurl{https://arxiv.org/abs/2606.31149}
\bibitem{Ye2024EMExposure}Z. Zhang, \emph{et al.}, ``Jointly correlated dual-side fluid antenna system,'' \emph{IEEE Wireless Commun. Lett.}, early access, 2026, doi: 10.1109/LWC.2026.3710907.
\bibitem{Wu2026BLPA}Y. Wu, \emph{et al.}, ``Learned blockwise port activation for real-time beamforming in fluid antenna arrays,'' \emph{arXiv preprint}, 2026. [Online]. Available: \arxivurl{https://arxiv.org/abs/2607.25365}
\bibitem{Wagner2012RZF}S. Wagner, \emph{et al.}, ``Large system analysis of linear precoding in correlated MISO broadcast channels under limited feedback,'' \emph{IEEE Trans. Inf. Theory}, vol.~58, no.~7, pp.~4509--4537, Jul.~2012.
\bibitem{Jang2017GumbelSoftmax}E. Jang, \emph{et al.}, ``Categorical reparameterization with Gumbel-Softmax,'' in \emph{Proc. ICLR}, 2017.
\bibitem{Bengio2013STE}Y. Bengio, \emph{et al.}, ``Estimating or propagating gradients through stochastic neurons for conditional computation,'' \emph{arXiv preprint}, 2013. [Online]. Available: \arxivurl{https://arxiv.org/abs/1308.3432}
\end{thebibliography}
\end{document}